\documentclass[twocolumn,prb,eqsecnum,showpacs,preprintnumbers,amsmath,amssymb]{revtex4}

\usepackage{epsfig,psfrag}
\usepackage{graphicx}
\usepackage{dcolumn}
\usepackage{bm}
\newcommand{\nz}{\mathbb{N}} 
\include{pdfsync}

\begin{document}
\title{ Landau levels, edge states, and strained magnetic waveguides
in graphene monolayers with enhanced spin-orbit interaction }
\author{Alessandro De Martino,$^1$ Artur H\"utten,$^2$ 
and Reinhold Egger$^2$ }
\affiliation{
$^1$~Institut f\"ur Theoretische Physik, Universit\"at zu K\"oln,
D-50937 K\"oln, Germany \\
$^2$~Institut f\"ur Theoretische Physik,
Heinrich-Heine-Universit\"at, D-40225 D\"usseldorf, Germany }

\date{\today}
\begin{abstract}
The electronic properties of a graphene monolayer in a
magnetic and a strain-induced pseudo-magnetic field are studied  
in the presence of  spin-orbit interactions (SOI) that are artificially enhanced, 
e.g., by suitable adatom deposition. For the homogeneous case, we 
provide analytical results for the Landau level eigenstates
for arbitrary intrinsic and Rashba SOI, including also the Zeeman field.
The edge states in a semi-infinite geometry are studied in the absence
of the Rashba term.  For a critical value of the magnetic field,
we find a quantum phase transition separating two phases
with spin-filtered helical edge states at the Dirac point.
These phases have opposite spin current direction.
We also discuss strained magnetic waveguides with inhomogeneous
field profiles that allow for chiral snake orbits.  Such waveguides
are practically immune to disorder-induced backscattering, and
the SOI provides non-trivial spin texture to these modes.
\end{abstract}


\maketitle
\section{Introduction}

The physics of graphene monolayers continues to attract 
a lot of attention and to provide a rich source of interesting
phenomena.\cite{review1,beenakker,review2}  By studying the effects
of the spin-orbit interaction (SOI) in a graphene layer,
where symmetry allows for an ``intrinsic'' ($\Delta$) and 
a ``Rashba'' ($\lambda$) term in the SOI,
Kane and Mele\cite{kane} made a remarkable discovery that  sparked
the exciting field of topological insulators:\cite{hasan}
For $\Delta>\lambda/2$, there is a bulk gap with topologically protected
edge states near the boundary of the sample. 
This is similar to the quantum Hall (QH) effect but happens in a time-reversal
invariant system. The resulting ``quantum spin Hall'' (QSH)
edge states form a one-dimensional (1D) helical liquid, where 
right- and left-movers have opposite spin polarization and 
spin-independent impurity backscattering is strongly suppressed.
The QSH state has been observed in HgTe quantum wells,\cite{qsh} but
several works\cite{huertas06,min,yao} showed that $\Delta$ is probably 
too small to allow for the experimental verification of this novel phase of 
matter in pristine graphene.  
Consequently, other material classes have 
been employed to demonstrate that topologically insulating behavior is 
indeed possible.\cite{hasan}
However, graphene experiments\cite{dedkov,vary} have also demonstrated that 
the Rashba coupling $\lambda$ can be increased significantly by depositing
graphene on Ni surfaces.  Moreover, 
very recent theoretical predictions\cite{franz}
suggest that already moderate indium or thallium adatom deposition will
dramatically enhance $\Delta$ by several orders of magnitude.   
By using suitable adatoms, it is expected that  in the near future
both SOI parameters $\Delta$ and $\lambda$ can be varied 
over a wide range in experimentally accessible  setups.

In view of these developments, in this paper we study the electronic
properties of a graphene monolayer with artificially enhanced SOI.   
Besides the SOI, we include piecewise constant electrostatic potentials, 
orbital and Zeeman magnetic fields, and
strain-induced vector potentials. The latter cause pseudo-magnetic fields 
but do not violate time reversal invariance; for a review, 
see Ref.~\onlinecite{strain}.  While the interplay of the Rashba 
term $\lambda$ with (pseudo-)magnetic fields in graphene
has been studied in several theory works before,\cite{castro09,huertas09,rashba}
the intrinsic SOI $\Delta$ did not receive much attention so far.
However, the transmission properties of  graphene's Dirac-Weyl (DW) 
quasiparticles through barriers with arbitrary SOI have been studied 
recently\cite{dario1,dario2} in the absence of (pseudo-)magnetic fields.

The structure of this article is as follows.  In Sec.~\ref{sec2} 
we formulate the model and construct the general solution for
piecewise constant fields.  On top of the orbital magnetic field, 
we allow for arbitrary SOI parameters $\Delta$ and  $\lambda$,
Zeeman energy $b$, and we also take into account aspects of strain-induced
fields.  The homogeneous case is addressed in Sec.~\ref{sec3}, where we
determine the Landau level states for this problem in closed and explicit form.
In particular, the fate of the zero modes residing at the Dirac point 
(energy $E=0$) will be discussed in the presence of the SOI. 
Our results also apply to the case of a
strain-induced homogeneous pseudo-magnetic
field.\cite{guineanp}
Next, in Sec.~\ref{sec4} we study edge states near the boundary
of a semi-infinite sample for vanishing Rashba coupling, $\lambda=0$. 
For weak magnetic fields, one then expects to have helical 
(spin-filtered) QSH edge states. Interestingly, at the Dirac point,
upon increasing the magnetic field, we find that a quantum
phase transition takes place between the QSH phase and a second QSH-like
phase with spin-filtered edge states, 
considered previously by Abanin \textit{et al.},\cite{abanin}
where the spin current direction is reversed.  This spin current
reversal should allow for an experimental detection of this
quantum phase transition, on top of the obvious consequences for 
QH quantization rules.\cite{gus,fertig1,abanin,peres}
In Sec.~\ref{sec5}, we turn to a mesoscopic waveguide geometry,
where a suitable inhomogeneous magnetic field (or exchange field
produced by lithographically deposited ferromagnetic films) defines the 
waveguide.\cite{ademarti,masir1,lambert,ghosh,haus1,luca1,luca2,ssc,heinz,luca3}  
We show that the SOI parameters $\Delta$ and $\lambda$ 
give rise to interesting spin texture of the resulting propagating
chiral states in such a waveguide. 
Finally, we conclude in Sec.~\ref{sec6}.

\section{Model and general solution} \label{sec2}

\subsection{Model}
\label{sec2a}

Unless many-body effects are of crucial importance, the low-energy
electronic properties of a graphene monolayer are well
captured by two copies of a DW Hamiltonian supplemented with various 
terms describing SOI, (pseudo-)magnetic fields, and electrostatic
potentials.\cite{review2}  The wavefunction corresponds to
a spinor comprising eight components, 
\begin{equation}\label{spinor}
\Psi(x,y) = \left( \begin{array}{c} 
\Psi_{A \uparrow K} \\ \Psi_{B \uparrow K} \\
\Psi_{A \downarrow K}\\ \Psi_{B \downarrow K} \\
\Psi_{A \uparrow K'}\\ \Psi_{B \uparrow K'} \\
\Psi_{A \downarrow K'}\\ \Psi_{B \downarrow K'} \end{array}\right)(x,y)=
e^{ik_x x} 
\left(\begin{array}{c} \phi^K (y) \\ \phi^{K'}(y) \end{array}
\right).
\end{equation}
The Pauli matrices $\sigma_{i=x,y,z}$ below act in 
sublattice space corresponding to the two carbon atoms ($A/B$) in the basis
of the honeycomb lattice, while Pauli matrices $s_i$
act in physical spin ($\uparrow,\downarrow$) space.  
Finally, the valley degree of freedom ($K,K'$) corresponds to the two 
$K$ points\cite{review2} and Pauli matrices $\tau_i$ refer to that space.
Specifically, we here consider models where the mentioned extra terms 
in the Hamiltonian are piecewise constant along the $y$-direction and
homogeneous along the $x$-axis.  Consequently, the momentum 
$p_x$ is conserved, and we have an 
effectively 1D problem in terms of the four-spinors $\phi^{K,K'}(y)$.
The orbital magnetic field 
$B_z=\epsilon B$ (with $\epsilon=\pm$ and $B\ge 0$) is expressed in 
terms of the vector potential ${\bf A}(x,y)$, 
where we choose the gauge 
\begin{equation}\label{vectorpotential}
A_x = - \epsilon B(y-c_0),\quad A_y=0.
\end{equation} 
Inclusion of the constant $c_0$ is necessary when connecting 
regions with different magnetic fields in order to make $A_x$ continuous.
Assuming that the magnetic field is perpendicular to the graphene sheet,
the Zeeman field couples to $s_z$ and determines the coupling 
constant $b= g_s \mu_B B/2$ 
, where 
$g_s\approx 2$ is the Land{\'e} factor
and $\mu_B$ denotes the Bohr magneton. 
The full Hamiltonian then reads\cite{review2} ($e>0$)
\begin{eqnarray}\nonumber
H &=& v_F \left[ \sigma_x\tau_z \left( p_x + \frac{e}{c}\left(
A_x +\tau_z {\cal A}_x \right) \right)
+  \sigma_y \left( p_y+\frac{e}{c} \tau_z {\cal A}_y \right) \right] \\ \label{ham}
& +& V+ \epsilon b s_z +
\frac{\lambda}{2} (\sigma_x s_y \tau_z -\sigma_y s_x)+ \Delta \sigma_z s_z 
\tau_z.
\end{eqnarray}
In Eq. (\ref{ham}) $p_x=\hbar k_x$ is the conserved 
momentum in the $x$-direction, 
while $p_y=-i \hbar \partial_y$ is still an operator.
The constant $c_0$ in Eq.~\eqref{vectorpotential} can be included by 
shifting $p_x$, and we suppose that this shift has been carried out 
in the remainder of this section. 
The Fermi velocity is $v_F\approx 10^6$~m$/$s, while the SOI couplings
$\Delta$ and $\lambda$ (both are assumed non-negative)
correspond to the intrinsic and Rashba terms, respectively.
In wrinkled graphene sheets the coupling $\lambda$ 
also captures curvature effects.\cite{huertas06}  A constant electrostatic 
potential, $V$, has been included in Eq.~\eqref{ham}. 
Strain-induced forces\cite{strain} lead to a renormalization of $V$ as 
well as to the appearance of an effective vector potential,
\[
\left( \begin{array}{c} {\cal A}_x \\ {\cal A}_y \end{array}\right)
= \kappa \left(\begin{array}{c} u_{xx}-u_{yy} \\-2 u_{xy} \end{array}\right),
\]
expressed in terms of the in-plane strain tensor $u_{ij}$, see
Ref.~\onlinecite{landau}.The constant $\kappa$ can be found in 
Refs.~\onlinecite{strain,ando}.
As discussed by Fogler {\it et al.},\cite{fogler} in many cases
it is sufficient to consider a piecewise constant strain
configuration.  Assuming that the
$x$-axis is oriented along the zig-zag direction,
strain causes only a finite but constant ${\cal A}_x$  while ${\cal A}_y=0$.
This can be taken into account by simply 
shifting $p_x$ in this region.  Below we suppose that also this shift
has already been done.  Estimates for ${\cal A}_x$ in terms of physical 
quantities can be found in Refs.~\onlinecite{strain,fogler}.
The resulting pseudo-magnetic field then consists of $\delta$-barriers
at the interfaces between regions of different strain.
An alternative situation captured by our model is given by
a constant pseudo-magnetic field, whose practical realization has
been described recently.\cite{guineanp} 
In that case, ${\cal A}_x$ is formally identical to $A_x$ in 
Eq.~\eqref{vectorpotential}.  Unless specified explicitly,
we consider the case of constant ${\cal A}_x$ below.

\subsection{Symmetries}
\label{sec2aa}

Let us briefly comment on the symmetries of this Hamiltonian.
In position representation, the time reversal transformation
is effected by the antiunitary operator\cite{footnote1}
\begin{equation}
{\cal T} = \tau_x (-is_y) {\cal C}
\end{equation}
with complex conjugation operator ${\cal C}$ and implies the relation
\begin{equation}\label{trs1}
{\cal T} H_\epsilon (k_x) {\cal T}^{-1} = H_{-\epsilon} (-k_x)
\end{equation}
for $H$ in Eq.~\eqref{ham} with $\epsilon={\rm sgn}(B_z)$. Since
$H$ is diagonal in valley space, Eq.~(\ref{trs1}) implies that 
the Hamiltonian $H^{K'}$ near the $K'$ point is related to $H^K$
by the relation
\begin{equation}\label{switch}
H^{K'}_{-\epsilon} (-k_x)= s_y [ H^K_\epsilon(k_x)]^* s_y.
\end{equation}
By solving the eigenvalue problem at the $K$ point, we could thus 
obtain the eigenstates at $K'$ via Eq.~\eqref{switch}.
A simpler way to achieve this goal is sketched at the end of this subsection.

From now on we switch to dimensionless
quantities by measuring all energies in units of the cyclotron energy
$\hbar\omega_c$, where we define $\omega_c=  v_F/\ell_B$.
The magnetic length $\ell_B=(\hbar c/2eB)^{1/2}$ sets the unit of length.
A field of 1 Tesla corresponds to $\hbar\omega_c\approx 36$~meV and
$\ell_B \approx 18$~nm.
Measuring $B$ in units of Tesla, we get for the
Zeeman coupling 
$b = (g_s \mu_B B /2)/\hbar \omega_c \approx 1.6\times 10^{-3} \sqrt{ B[{\rm T}]}$.
With the dimensionless coordinate 
\begin{equation}\label{etadef}
\eta=y-2\epsilon k_x
\end{equation}
and the auxiliary quantities
\begin{equation} \label{mudef}
\mu_\pm = E-V + b \pm \Delta,\quad \nu_\pm =E-V-b\pm \Delta,
\end{equation}
we find the representation
\begin{eqnarray}\label{det1}
E-H^K_{\epsilon=+1} &=& \left( \begin{array}{cccc}
\nu_- & a & 0 & 0 \\
a^\dagger & \nu_+ & i\lambda & 0\\
0 & -i\lambda & \mu_+ &  a\\
0 & 0  &  a^\dagger & \mu_-
\end{array}\right) ,\\ \nonumber
E-H^K_{\epsilon=-1} &=& \left( \begin{array}{cccc}
\mu_- & -a^\dagger & 0 & 0 \\
-a & \mu_+ & i\lambda & 0\\
0 & -i\lambda & \nu_+ & - a^\dagger\\
0 & 0  & - a & \nu_-
\end{array}\right).
\end{eqnarray}
Here we introduced the standard ladder operators 
\begin{equation}\label{adef}
a =  \frac{\eta}{2} + \partial_\eta,\quad 
a^\dagger= \frac{\eta}{2}-\partial_\eta,
\end{equation}
with $[a,a^\dagger]=1$.

According to the above discussion,
eigenstates at the $K'$ point for $\epsilon=\pm 1$ 
could be obtained from the corresponding solutions at the $K$ point
with $\epsilon=\mp 1$. Alternatively, there is a simpler way to obtain
the $K'$ states as follows. The 1D Hamiltonians
$H^{K,K'}$ (for given $\epsilon$) can be written in dimensionless notation as
\begin{eqnarray*}
H^K &=& -\frac{\epsilon \eta}{2} \sigma_x -i\sigma_y\partial_\eta +
\Delta \sigma_z s_z +\\
&+& \frac{\lambda}{2}(\sigma_x s_y-\sigma_y s_x)
+ {\cal A}_x \sigma_x +\epsilon b s_z ,\\
H^{K'} &=& \frac{\epsilon \eta}{2} \sigma_x -i\sigma_y\partial_\eta -
\Delta \sigma_z s_z +\\
&+& \frac{\lambda}{2}(-\sigma_x s_y-\sigma_y s_x)
+ {\cal A}_x \sigma_x +\epsilon b s_z.
\end{eqnarray*}
Both Hamiltonians are therefore related by the transformation
\begin{equation}
H^{K'}({\cal A}_x) = \sigma_y H^K(-{\cal A}_x) \sigma_y,
\end{equation}
without the need to invert the real magnetic field since this is
not a time reversal transformation.  As a consequence, the 1D eigenstates
$\phi^{K'}(\eta)$ follow from the solutions at the $K$ point by multiplying
with $-i\sigma_y$ and inverting the sign of ${\cal A }_x$,
\begin{equation}\label{maptokprime}
\phi^{K'}(\eta,{\cal A}_x)  =  - i\sigma_y \phi^K(\eta,-{\cal A}_x).
\end{equation}

\subsection{General solution}
\label{sec2b}

We now determine the spinors $\phi$ solving the DW equation for energy $E$,
\begin{equation} \label{schr}
(E-H^K)\phi(\eta)=0,
\end{equation}
with $E-H^K$ in Eq.~\eqref{det1}.  We construct the solution to 
Eq.~\eqref{schr} within a spatial region where 
all parameters (magnetic fields, strain, SOI, etc.) are constant but
arbitrary.  This general solution will be employed in later sections, where 
specific geometries are considered by matching wavefunctions 
in adjacent parts.  Now Eq.~\eqref{schr} is a system of four coupled 
linear differential equations that admits precisely
four linearly independent solutions derived in App.~\ref{appa}.  
In order to solve Eq.~\eqref{schr}, it is instructive to realize that the
parabolic cylinder functions,\cite{abram,grad} $D_p(z)$, obey the recurrence
relations 
\begin{equation} \label{parbol}
a D_p(\eta) = p D_{p-1}(\eta), \quad a^\dagger D_p(\eta) = D_{p+1}(\eta),
\end{equation}
with the ladder operators $a,a^\dagger$ in Eq.~\eqref{adef}.   
Similar relations for $\eta\to -\eta$ or $\eta\to i\eta$ are given in 
App.~\ref{appa}.  For given energy $E$, 
the order $p$ can only take one of the two values 
\begin{equation}\label{pdef}
p = \frac12 \left[ \mu+\nu-1\pm 
\sqrt{(\mu+\nu-1)^2+4\lambda^2\mu_-\nu_-}\right],
\end{equation}
where we define [cf.~Eq.~\eqref{mudef}]
\begin{eqnarray}\label{aux2}
\mu&=&\mu_+\mu_-= (E-V+b)^2-\Delta^2,\\ \nonumber
\nu&=&\nu_+\nu_-= (E-V-b)^2-\Delta^2.
\end{eqnarray} 
For each of the two possible values for $p$, we then have two 
basis states, $\phi_p$ and $\psi_p$, which results in 
four linearly independent solutions.
We show in App.~\ref{appa} that the (unnormalized) solution
$\phi_p$ can be chosen as
\begin{eqnarray}\label{sol1}
\phi_{\epsilon=+1,p} &=&\left( \begin{array}{c} 
p D_{p-1}(-\eta) \\ \nu_- D_p(-\eta)\\
\frac{i(\nu-p)}{\lambda} D_p(-\eta)\\ \frac{ i(\nu-p)}{\lambda\mu_-}
D_{p+1}(-\eta) \end{array} 
\right), \\ \nonumber \phi_{\epsilon=-1,p} & =& 
\left( \begin{array}{c} 
D_{p+1}(\eta) \\ \mu_- D_p(\eta)  \\
\frac{i(\mu-p-1)}{\lambda} D_p(\eta)\\
\frac{i(\mu-p-1)}{\lambda\mu_-} pD_{p-1}(\eta) \end{array} \right),
\end{eqnarray}
while $\psi_p$ is taken in the form
\begin{eqnarray}\label{sol2}
\psi_{\epsilon=+1,p} & = & \left( \begin{array}{c}
- i D_{-p}(-i\eta) \\  \nu_- D_{-p-1}(-i\eta)\\ 
\frac{i(\nu-p)}{\lambda} D_{-p-1}(-i\eta)\\
-\frac{(\nu-p)(p+1)}{\lambda\mu_-}
D_{-p-2}(-i\eta) \end{array} \right),
\\ \nonumber
\psi_{\epsilon=-1,p} & =&
\left( 
\begin{array}{c}  i (p+1)D_{-p-2}(i\eta) \\
\mu_- D_{-p-1}(i\eta)\\
\frac{i(\mu-p-1)}{\lambda} D_{-p-1}(i\eta)\\
\frac{\mu-p-1}{\lambda\mu_-} D_{-p}(i\eta)
\end{array} \right).
\end{eqnarray}
Next, we analyze the spatially homogeneous case.

\section{Homogeneous case}
\label{sec3}

In this section we study an unstrained 
infinitely extended graphene monolayer where 
the magnetic field $B_z=B$ (we assume $\epsilon=+1$)
and the SOI parameters $\Delta$ and $\lambda$ are constant everywhere.
(The electrostatic potential $V$ just shifts all states and is set
to zero here.) We are thus concerned with the 
relativistic Landau level structure for graphene in the presence
of arbitrary SOI parameters, including also the Zeeman field $b$.   
This problem was solved for the special case $\Delta=b=0$ 
by Rashba,\cite{rashba} see also Ref.~\onlinecite{huertas09}, 
and below we reproduce and generalize this solution.
We focus on the $K$ point only, since 
the spectrum and the eigenstates at  the $K'$ point 
follow from Eqs.~\eqref{switch}  and \eqref{maptokprime}.
We also allow for a constant pseudo-magnetic field.
When only an orbital or a strain-induced pseudo-magnetic
field is present but not both, each energy level below has
an additional twofold valley degeneracy.

In the homogeneous case, normalizability of the spinors $\phi_p$ 
[Eq.~\eqref{sol1}] can only be satisfied if the order $p$
is constrained to integer values $p=-1,0,1,2,\ldots$, while
the $\psi_p$ [Eq.~\eqref{sol2}] are not normalizable. Solutions for the
homogeneous problem thus have to be constructed using $\phi_p$ only.
Expressing the energy $E$ (we remind the reader that here all
energy scales are measured in units of $\hbar \omega_c$)
in terms of $p$ [Eq.~\eqref{pdef}],
the sought (valley-degenerate) Landau levels 
follow as the roots of the quartic equation
\begin{eqnarray}\nonumber && 
\left[ (E+b)^2 - (p+1+\Delta^2) \right] \left[(E-b)^2-(p+\Delta^2)\right]=
\\ && = \label{quartic}
\lambda^2 \left[ (E-\Delta)^2-b^2\right].
\end{eqnarray}
For $b=\lambda=\Delta=0$ this recovers the standard relativistic
spin-degenerate Landau levels,\cite{review2}
$E_{\pm,n}=\pm \sqrt{n}$ for $n=1,2,3,\ldots$ (with 
$n=p$ for spin up and $n=p+1$ for spin down states), 
plus  a spin-degenerate zero mode $E_0=0$ (for $p=0,-1$).
We notice from Eq.~\eqref{quartic} that for $b=0$,
the combination of $\Delta$ and $\lambda$  breaks
particle-hole symmetry, while the two couplings individually keep
it.  Furthermore, zero-energy solutions  are generally not possible
except for special fine-tuned parameters.  Eq.~\eqref{quartic}
also predicts that if $E$ is a solution for the parameter set
$\{p,\lambda,\Delta,b\}$ then $-E$ is a solution for the set $\{p,\lambda,-\Delta,-b\}$. 
The $\phi_p(\eta)$ thus represent Landau level states 
in the presence of SOI and Zeeman coupling.  
The normalization constant $1/\sqrt{{\cal N}_p}$,
entering as a prefactor in Eq.~\eqref{sol1}, 
can be computed analytically since  $D_p(z)$ can be expressed in terms
of Hermite functions for integer $p$.\cite{grad}  
For $p=1,2,3,\ldots$, we find
\begin{eqnarray}\label{norm}
{\cal N}_p &=& \frac{\sqrt{2\pi} \ p !}{(\lambda\mu_-)^2}
\Bigl[ (\lambda\mu_-)^2p + \\ \nonumber && \quad +
\mu_-^2(\lambda^2\nu_-^2+(\nu-p)^2)+ (\nu-p)^2(p+1)\Bigr]. 
\end{eqnarray}
Remarkably, for $p=-1$, we find the \textit{exact}\ normalized 
state for arbitrary system parameters, 
\begin{equation}\label{p-1}
\phi_{-1}(\eta) = \frac{1}{(2\pi)^{1/4}}
\left(\begin{array}{c} 0\\ 0\\ 0\\ D_0(-\eta) 
\end{array}\right),
\end{equation}
with the eigenvalue 
\begin{equation}\label{ep-1}
E_{p=-1}=\Delta-b.
\end{equation}
This unique admissible eigenstate for $p=-1$ is 
endowed with full spin polarization  in the $\downarrow$ direction.
For $p=0$, 
the secular equation (\ref{quartic})
becomes effectively a cubic equation: the solution
$E=\Delta+b$ (i.e., $\nu_-=0$)
does not correspond to any admissible eigenstate. 
The three allowed states are described by 
\begin{eqnarray}\label{p0}
\phi_{p=0}(\eta) &=& \frac{1}{\sqrt{{\cal N}_0}} 
\left(\begin{array}{c} 0 \\ \lambda \mu_- \nu_- D_0(-\eta)\\
i\mu_- \nu D_0(-\eta)\\ i\nu D_1(-\eta)\end{array}\right),
\\ \nonumber {\cal N}_0 &=& \sqrt{2\pi} 
\left[\nu^2(1+\mu_-^2)+\lambda^2\mu_-^2 \nu^2_-\right].
\end{eqnarray}
This includes a ``zero-mode'' partner of 
the $p=-1$ state, plus a pair of states obtained by mixing the 
spin-up $n=0$ and spin-down $n=\pm 1$ Landau orbitals via the Rashba SOI.

\subsection{Rashba SOI only}\label{sec3r}

For $\Delta=b=0$ but allowing for a finite Rashba SOI 
parameter $\lambda$, Eq.~\eqref{quartic} admits a simple solution,
previously given in Ref.~\onlinecite{rashba} and briefly summarized here
for completeness.
For $p=-1$ we have the solution (\ref{p-1}), which now is
a zero mode, while for $p=0,1,2,\ldots$, the eigenenergies are given by
\begin{eqnarray}\label{rashba}
E_{p,\alpha,\beta} &=& \alpha  \Biggl[
\frac{1+\lambda^2}{2} + p + \\ \nonumber &+& \beta  \sqrt{\left(\frac{1
+\lambda^2}{2} +p\right)^2-p(p+1)} \  \Biggr]^{1/2},
\end{eqnarray}
with $\alpha,\beta=\pm$.
According to our discussion above, 
here $E_{0,\pm,-}=0$ should be counted only once, with eigenstate
$\phi_{0,\;,-}^T\propto (0,D_0(-\eta),0,-i\lambda D_1(-\eta))$,
while $E_{0,\pm,+}=\pm \sqrt{1+\lambda^2}$ 
correspond to a particle/hole pair of first Landau levels modified by 
the Rashba SOI, with eigenstates $\phi_{0,\pm,+}^T \propto 
(0, \lambda D_0(-\eta), \pm i\sqrt{1+\lambda^2}D_0(-\eta), i D_1(-\eta))$.
We thus get precisely two zero-energy states.

For small $\lambda$, we find  the expansion
\begin{eqnarray*}
E_{p-1,\pm,+}&=&\pm (1+\lambda^2/2)\sqrt{p} + {\cal O}(\lambda^4),\\
E_{p,\pm,-}&=&\pm (1-\lambda^2/2)\sqrt{p} + {\cal O}(\lambda^4),
\end{eqnarray*}
which shows that the states $E_{p,\pm,+}$ and $E_{p+1,\pm,-}$, which form
a degenerate Landau level for $\lambda=0$, are split by a finite $\lambda$.

\subsection{Intrinsic SOI only}\label{sec3i}

Let us next consider the case $\lambda=0$, where one has a
QSH phase\cite{kane} for $B=0$ and $\Delta\ne 0$. Now
the Hamiltonian is block diagonal in spin space and the eigenstates 
become quite simple even for finite Zeeman coupling, 
since we can effectively work with the bi-spinors
$\phi^{K,K'}_{\uparrow, \downarrow}(y)$  for 
spin $s=\uparrow/\downarrow=\pm$.
We easily obtain the (unnormalized) eigenstates with 
$p\in \mathbb{N}_0$ in the form\cite{foot5}
\begin{eqnarray}\label{e4ed}
\phi^K_{p,\pm,s} (\eta) &= & \left( \begin{array}{c} 
\nu_{p,\pm,s} D_{p-1}(-\eta) \\ D_p(-\eta) \end{array} \right)
, \\ \nonumber
\phi^{K'}_{p,\pm,s} (y) &= & \left( \begin{array}{c}
-D_p(-\eta) \\  \nu_{p,\pm,s} D_{p-1}(-\eta)  \end{array} \right),
\end{eqnarray}
where the eigenenergies follow from Eq.~\eqref{quartic},
\begin{equation}\label{soir}
E_{p,\pm,s}= s b \pm \sqrt{p+\Delta^2}.
\end{equation}
We employ the notation
\begin{equation}\label{nudef}
\nu_{p,\pm,s} \equiv E_{p,\pm,s}-E_{0,-s,s} =\pm \sqrt{p+\Delta^2} -s\Delta.
\end{equation}
For $p=0$, the second index in $\phi_{p,\pm,s}$ and $E_{p,\pm,s}$
should be replaced by $ -s$, i.e., there is only one
solution for given spin (and valley). 
Note that $E_{0,+,\downarrow}$ in the present notation
corresponds\cite{foot5} to the solution \eqref{p-1}.  
When $b=0$, interestingly enough, 
$\Delta$ does \textit{not}\ lift the spin degeneracy of the Landau 
levels except for the zero mode ($p=0$).\cite{foot2}
A Zeeman term with $b=\Delta$ restores a true doubly-degenerate
zero-energy state for $p=0$ again. In Sec.~\ref{sec4} we show
that this implies a quantum phase transition.

\subsection{General case}
\label{sec3gen}

Although the quartic equation (\ref{quartic}) can be solved analytically 
when both SOI couplings are finite, the resulting expressions are 
not illuminating and too lengthy to be quoted here.  
Only the $p=-1$ state in Eq.~\eqref{p-1} remains  
exact for arbitrary parameters. 
We here specify the leading perturbative
corrections around the special cases above, and then show the
generic behavior in two figures. 

Expanding around the Rashba limit of Sec.~\ref{sec3r}, which is
justified for $b,\Delta\ll 1 $, we get
the lowest-order perturbative correction to 
the finite-energy (i.e., $p\neq 0,-1$)
Landau levels (\ref{rashba}) in the form
\begin{equation}
\delta E_{p,\pm,+}=-\delta E_{p,\pm,-}= \frac{(\lambda^2\Delta+b)}
{\sqrt{(1+\lambda^2)^2 +4p\lambda^2 }}.
\end{equation}

Expanding instead around the intrinsic SOI limit of Sec.~\ref{sec3i}, 
we find the following small-$\lambda$ corrections to the 
Landau levels  in Eq.~\eqref{soir}:\cite{foot5}  For $p=0$, the state $E_{0,+,\downarrow}$
corresponding to the exact solution (\ref{p-1}) 
is not changed by $\lambda$ to any order,
while $E_{0,-,\uparrow}$ obtains the lowest-order correction 
\[
\delta E_{0,-,\uparrow}= \frac{ 2(\Delta-b)\lambda^2}{4b(b-\Delta)+1} .
\]
The corresponding eigenstate is, however,  not a spin-$\uparrow$ state anymore.
For $p> 0$, the eigenenergy $E_{p,\pm,s}$ 
[Eq.~\eqref{soir}] acquires the perturbative correction
\begin{widetext}
\begin{equation}
\delta E_{p,\pm,s} = \pm \frac{s\lambda^2}{2\sqrt{p+\Delta^2}}
\frac{p+2(\Delta-s b)(\Delta\mp \sqrt{p+\Delta^2})}
{1+4b\left(s b\pm \sqrt{p+\Delta^2}\right) }.
\end{equation}
\end{widetext}

\begin{figure}
\scalebox{0.3}{\includegraphics{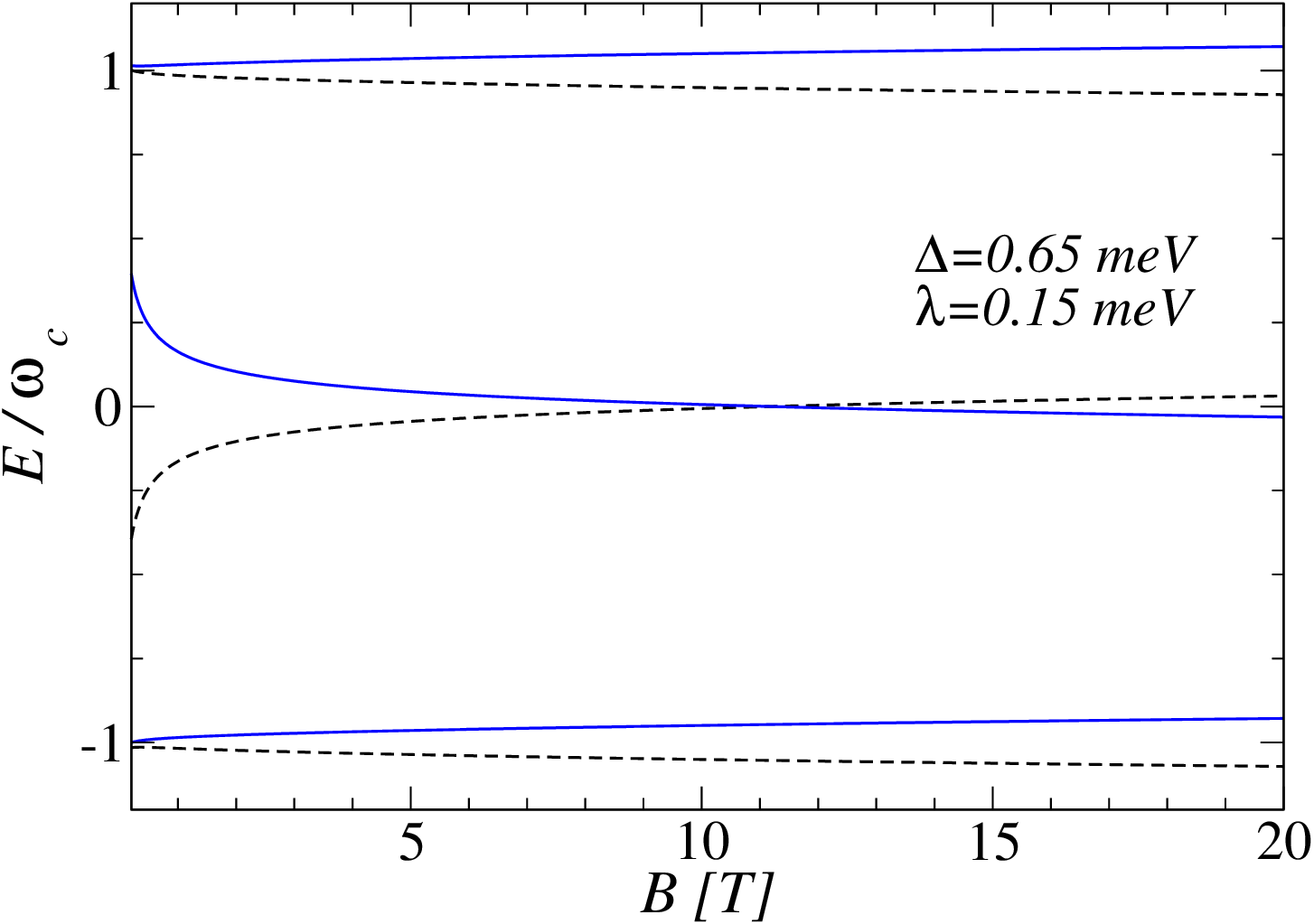}}
\caption{\label{fig1} (Color online)
Low-lying Landau level energies (in units of 
the cyclotron energy $\hbar \omega_c$) vs magnetic field $B$ (in Tesla) 
for the SOI parameters 
$\Delta=0.65$~meV and $\lambda=0.15$~meV. 
For small $B$, this corresponds to the QSH phase, $\Delta>\lambda/2$. 
For better visibility, the deviation from the respective
$\Delta=\lambda=b=0$ level has been magnified by a factor 10
for each curve. }
\end{figure}

\begin{figure}
\vspace{2cm}
\scalebox{0.3}{\includegraphics{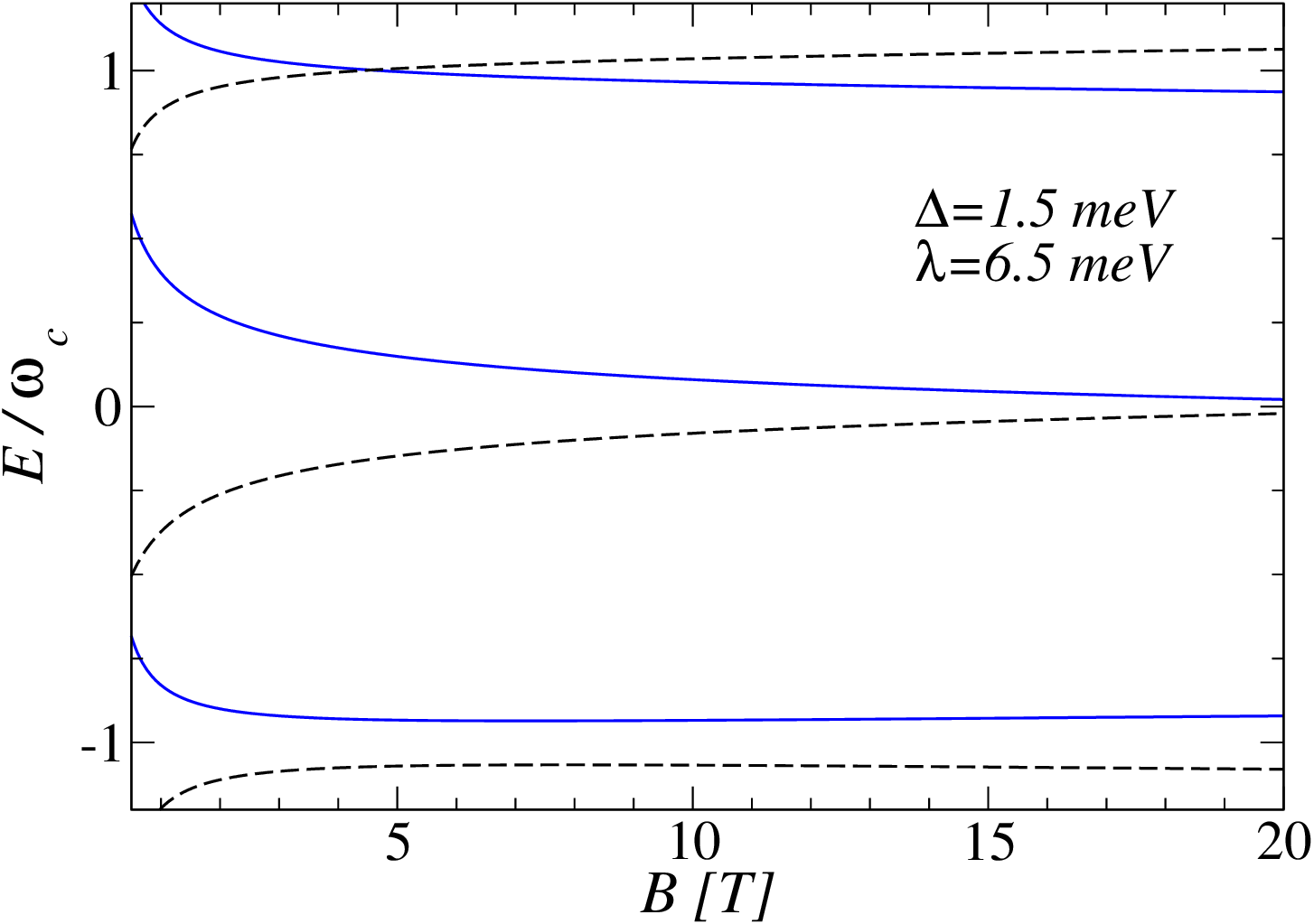}}
\caption{\label{fig2} (Color online) Same as in Fig.~{\ref{fig1}} but for  
$\Delta=1.5$~meV and $\lambda=6.5$~meV.
}
\end{figure}

We now consider two different SOI parameter sets 
consistent with the estimates in Ref.~\onlinecite{franz},
and show the complete evolution of the Landau levels from the
weak- to the strong-field limit.  In Fig.~\ref{fig1}, numerical
results for the few lowest-energy Landau levels 
are depicted for $\Delta>\lambda/2$, corresponding to a QSH 
phase for $B=0$. The  (valley-degenerate) spin-split levels
corresponding to the $\Delta=\lambda=b=0$ zero mode
exhibit a zero-energy crossing at $B\approx 11$~T for the chosen 
SOI parameters. This crossing signals a 
quantum phase transition from the QSH phase, which survives for
sufficiently small $B$ and $\Delta>\lambda/2$, to a peculiar QH phase
for large $B$.
As we discuss in Sec.~\ref{sec4}, 
one then again has helical edge states\cite{abanin} but 
with reversed spin current.
Similar crossings can occur for higher Landau states as well, as is shown
in Fig.~\ref{fig2} for a parameter set with $\Delta<\lambda/2$ where
no QSH physics is expected.
For even larger $B$ not displayed in Fig.~\ref{fig2}, 
we find an $E=0$ crossing where the Rashba-dominated small-$B$ phase turns 
into the helical QH phase.

\subsection{Spin polarization}

Given the Landau level eigenstates,
it is straightforward to compute the spin-polarization densities
$S_i(y) = \Psi^\dagger \frac{s_i}{2} \Psi$ ($i=x,y,z$).
We find $S_x(y)=0$, while
\begin{eqnarray}\label{sy}
S_y(y) &=&  \frac{\nu-p}{\lambda {\cal N}_p}
\left( p D_{p-1} D_p + \frac{\nu_-}{\mu_-} D_p D_{p+1} \right),\\
S_z(y) &=& \frac{1}{2{\cal N}_p} \Biggl[ \nonumber
p^2 D_{p-1}^2+\left( \nu_-^2-\frac{(\nu-p)^2}{\lambda^2}\right) D_p^2 \\
&& -\quad\frac{(\nu-p)^2}{\lambda^2\mu_-^2} D_{p+1}^2 \Biggr] \nonumber,
\end{eqnarray}
where $D_p\equiv D_p(-\eta)$.
In the absence of the Rashba term ($\lambda=0)$,  the in-plane
component $ S_y $ vanishes identically, since then
the eigenstates are simultaneously eigenstates of  $s_z$. 
For finite $\lambda$, integration over $y$ yields
a vanishing expectation value for the overall in-plane polarization, 
but the Rashba coupling still induces {\it local}\ in-plane spin polarization.
The case $\Delta=b=0$ has been discussed in detail by Rashba.\cite{rashba}

\section{QH edge states for intrinsic SOI}
\label{sec4}

In this section, we consider the edge states corresponding to the 
relativistic Landau level problem in Sec.~\ref{sec3} when
a boundary at $y=0$ is present.  We focus on the case of 
purely intrinsic SOI, $\lambda=0$, but the physics should be 
qualitatively unchanged for $\lambda\ll \Delta$.
In the region $y<0$ we then have a homogeneous magnetic
field  $B_z=+B$, i.e, $\epsilon=+1$.  (For a pseudo-magnetic field,
this holds at the $K$ point while at the $K'$ point, $B_z \to -B_z$.)

Since the problem of edge states in graphene has been studied
extensively before, some remarks are in order at this point.
In fact, putting $\Delta=b=\lambda=0$, our results are consistent with those of 
Refs.~\onlinecite{fertig1,peres,kormanyos1,mont,landman} reporting chiral
QH edge states in graphene.  On the other hand, the $B=0$ model is 
equivalent to the continuum limit of the Kane-Mele model\cite{kane} and thus
exhibits helical QSH edge states.\cite{qsh} 
(The helical state has a pair of counterpropagating 1D modes
with opposite spin polarization.)  
The Kane-Mele model with $(\Delta,b)\ne 0$ but without orbital 
magnetic field has recently been studied,\cite{yang} 
and a quantum phase transition from a (generalized) QSH phase for $b<\Delta$
to a quantum anomalous Hall (QAH) phase for $b>\Delta$ has
been predicted.  It is worthwhile to stress that 
the QSH effect survives even when time-reversal symmetry is broken.
In the QAH phase, one has chiral edge states moving in the
same direction for both spin polarizations.\cite{alan} The 
valley analogue of this quantum phase transition has also been 
studied.\cite{niu} 
Furthermore, for the 2D topological insulator realized
in HgTe quantum well structures, a related transition has 
been predicted\cite{tkachov} by including the orbital field but
omitting the Zeeman term.

However, the Zeeman term is crucial in graphene near the 
Dirac point: for $\Delta=0$ and $b\ne 0$, 
spin-filtered helical edge states (similar to the QSH case) 
emerge again.\cite{abanin,arikawa}
Our results below show that this QSH-like phase is 
separated from the ``true'' QSH phase 
by a \textit{quantum phase transition}\ at $b=\Delta$.
Albeit both phases have spin-filtered edge states, they differ in
the \textit{direction}\ of the spin current. 
This feature should allow to experimentally
distinguish both phases and to identify the quantum phase
transition separating them.  In practice, one may reach this transition 
simply by changing the magnetic field.

Normalizability of the wavefunctions for $y\to -\infty$
implies\cite{abram} that the only allowed solutions 
follow from the $\phi_p$ spinors in Eq.~\eqref{sol1}, 
while the $\psi_p$ solutions [Eq.~\eqref{sol2}] have to be discarded.  
Since we do not have to impose normalizability 
at $y\to \infty$, the order $p$ is not constrained
to integer values and can now take any real value
consistent with suitable boundary conditions at $y=0$.  
For given conserved momentum $k_x$ and spin $s$, 
the solutions for $p$ yield the edge state spectrum, $E_s(k_x)$.
Note that for finite magnetic field and $k_x<0$, 
the distance from the boundary is set by $|k_x|$.
Putting $\lambda=0$, possible solutions $\phi^{K,K'}_{p,\pm,s}(y)$
must be of the form in Eq.~\eqref{e4ed}, with energy $E_{p,\pm,s}$ 
given by Eq.~\eqref{soir}.  While $p\in \mathbb{N}_0$ in Sec.~\ref{sec3i}, 
we now consider arbitrary real $p$.
To make progress, we have to specify boundary conditions at $y=0$.
We investigate two widely used boundary conditions, 
namely the zig-zag edge and the armchair edge.\cite{review2,abanin,fertig,akh}

\subsection{Zig-zag edge}

For a zig-zag edge with the last row of carbon atoms residing
on, say, sublattice  $A$,   the microscopic wavefunction must vanish 
on the next row outside the sample, belonging to sublattice $B$.
In the continuum limit, since the $x$-axis here points in the zig-zag
direction, the lower component of 
the spinor $\phi^K_{p,\pm,s}$ [Eq.~\eqref{e4ed}] has to vanish
at $y=0$.\cite{abanin,fertig1}
For both spin directions $s=\pm$, this yields the condition
\begin{equation}\label{c1}
D_p(2k_x)= 0,
\end{equation}
which has to be solved for the energy, expressed in terms of 
$p$ as $E_s=sb\pm\sqrt{p+\Delta^2}$.  At the other Dirac point,
the lower component of the spinor $\phi^{K'}_{p,\pm,s}$
should vanish at $y=0$, where Eq.~\eqref{maptokprime} implies the condition
\begin{equation}\label{c2}
\nu_{p,\pm,s} D_{p-1}(2k_x) = 0,
\end{equation}
with $\nu_{p,\pm,s}$ in Eq.~\eqref{nudef}.
It is not possible to find simultaneous solutions to both
Eqs.~\eqref{c1} and \eqref{c2}. Possible 
states are thus confined to a single valley: the boundary
condition does not mix the valleys but lifts the $KK'$ degeneracy.
Remarkably, for $s=\pm$ and arbitrary $k_x$, Eq.~\eqref{c2} 
is satisfied by the $K'$ solution for $p=0$ in Sec.~\ref{sec3i},
with $E_s(k_x) = s (b-\Delta)$, i.e., we find 
a pair of ``flat'' states.  
For all other states, Eq.~\eqref{c2} simplifies to condition (\ref{c1})
with $p\to p-1$ (and $K\to K'$).
We mention in passing that for $\Delta=0$ this condition
reduces to Eq.~(9) in Ref.~\onlinecite{landman}.
Equation (\ref{c1}) can be solved in closed form for $k_x\to -\infty$
using asymptotic properties of the parabolic cylinder function.
To exponential accuracy, with $n \in \nz_0$ we find
\begin{equation}
p = n + \frac{|2k_x|^{2n+1}}{\sqrt{2\pi} n!}  e^{-2k_x^2}.
\end{equation}
Numerical analysis of the above equations recovers 
the expected spin-filtered helical edge states\cite{abanin} for $b>\Delta$,
but the continuum approach used in this paper fails
to give clear evidence for the helical QSH edge states for $b<\Delta$. 
As pointed out in Ref.~\onlinecite{arikawa}, under the 
zig-zag boundary condition one needs a more microscopic description 
in order to capture these states. The ``flat'' states above
are remnants of the sought QSH edge states, but the continuum
model is not sufficient to describe their proper dispersion relation.  
We therefore turn to the armchair boundary condition.

\subsection{Armchair edge}

Under the armchair boundary condition, we instead
impose $\Psi_A^K+\Psi_A^{K'}=0$  and $\Psi_B^K+\Psi_B^{K'}=0$ 
at the boundary, with $\Psi$ in Eq.~\eqref{spinor}.
This boundary condition mixes the valleys and involves both 
sublattices.  Since in our coordinate system
the $x$-axis is parallel to the zig-zag
direction, we first rotate the system by $\pi/2$ and 
then impose the boundary condition at $y=0$.  Written in the original
coordinates, we find (for each spin direction $s$)
\begin{equation}\label{c3}
\nu_{p,\pm, s} D_{p-1}(2k_x) \pm D_p (2k_x) = 0.
\end{equation}
We note that the relative phase between the $K$ and $K'$ components
is not fixed by the Dirac equation, which is diagonal in valley space.
However, the only relative phase compatible with the boundary 
condition imposed simultaneously on both sublattices is $\pm 1$.
Each of the two conditions in Eq.~\eqref{c3} may thus
be imposed separately.  We have checked that the numerical 
solution of Eq.~\eqref{c3} for $\Delta=0$ recovers the known results
for the QH edge state spectrum.\cite{abanin,landman}
In addition, for $B=0$, the armchair edge is known \cite{fertig,sandler}
to yield QSH edge states.

\begin{figure}
\includegraphics[width=0.45\textwidth]{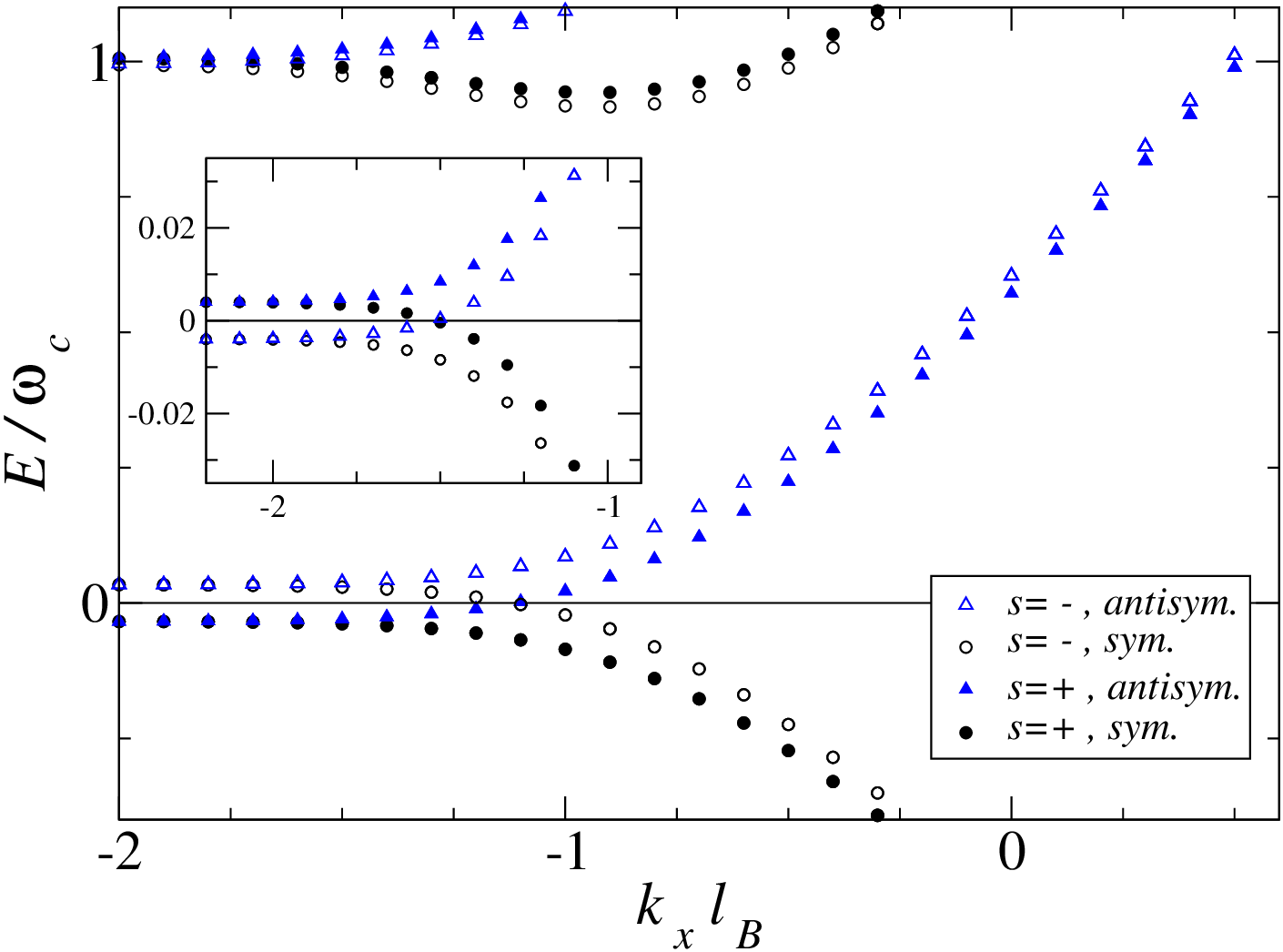}
\caption{\label{fig3}  (Color online)   
Dispersion relation $E_{s,\pm}(k_x)$ of a semi-infinite graphene sheet 
with an armchair edge at $y=0$,
obtained numerically from Eq.~\eqref{c3}. We use $\lambda=0$, $B=15$~T,
$\Delta=6$~meV, and the $+$ ($-$) sign is for the symmetric (antisymmetric)
valley combination in Eq.~\eqref{c3}.  Inset: Same for $\Delta=0.3$~meV.  }
\end{figure}

Our numerical results for the dispersion relation $E_{s,\pm}(k_x)$
for the armchair edge are shown in Fig.~\ref{fig3},
where $\pm$ corresponds to the symmetric or antisymmetric linear combination
in Eq.~\eqref{c3} and the magnetic field is $B=15$~T.
The main panel shows results for $\Delta=6$~meV. Then $\Delta>b$, 
and we have the (generalized) QSH phase.  
Indeed, for $E=0$ we find the helical edge state, where the 
right- (left-)mover has  spin $s=\uparrow$ $(s=\downarrow)$.
The inset of Fig.~\ref{fig3} is for $\Delta=0.3$~meV, 
where $\Delta<b$ and the spin-filtered helical QH phase\cite{abanin}
is found.  Here we have spin $s=\downarrow$ ($s=\uparrow$) for the 
right- (left-)mover. Hence the {\it spin current}\ 
differs in sign for $\Delta>b$ and $\Delta<b$, 
with a quantum phase transition at $\Delta=b$ separating both phases.
This feature should allow for an experimentally observable signature
of the transition.

\section{Spin structure in magnetic waveguides}\label{sec5}

In this section, a spatially inhomogeneous situation is considered,
where a magnetic waveguide\cite{lambert,ghosh,haus1} along the
$x$-direction can be realized. Since the problem remains homogeneous
along the $x$-direction, $p_x=\hbar k_x$ is still conserved.
For the physics described below,  the Zeeman coupling $b$
gives only tiny corrections\cite{haus1} and will be neglected.
Moreover, there are no valley-mixing terms such that we 
can focus on a single valley. 

We distinguish a central strip of width $2L$ (the ``waveguide"), 
$-L<y<L$, and two outer regions $y<-L$ and $y>L$.
In the central strip, we shall allow for arbitrary SOI parameters 
$\Delta$ and $\lambda$. In addition, strain may cause a constant 
contribution to the vector potential, ${\cal A}_x$, and a scalar 
potential, $V$.  The magnetic field in the central strip is 
denoted by $B_c$.
For $|y|>L$, we assume that all strain- or SOI-related effects
can be neglected, $\Delta=\lambda={\cal A}_x=V=0$.  
In principle, by lithographic deposition of adatoms, one may
realize this configuration experimentally. 
For $y<-L$, the magnetic field is $B_z=B>0$, while for 
$y>L$, we set $B_z=\epsilon B$, where $\epsilon=1$ ($\epsilon=-1$) 
corresponds to the parallel (antiparallel) field orientation on both
sides. For $\epsilon=-1$, we take $B_c=0$, while for $\epsilon=+1$,
we set $B_c=-B$.

The setup with $\epsilon=-1$ could be realized by
using a ``folded'' geometry,\cite{folded,rainis} 
cf.~recent experimental studies.\cite{fold2}
Note that when the magnetic field changes sign,
one encounters ``snake orbits,''  which
have been experimentally observed in graphene $pn$ junctions.\cite{snake}
For the $\epsilon=-1$  configuration, we have uni-directional 
snake orbits mainly localized along the waveguide, while for 
$\epsilon=+1$, we get two counterpropagating snake states 
centered near $y=\pm L$.  
For $\Delta=\lambda={\cal A}_x=V=0$, both cases ($\epsilon=\pm 1$) have
been studied in detail in Ref.~\onlinecite{ghosh}.  Technically,
one determines the eigenstates and the spectrum, $E(k_x)$,
by matching the wavefunctions in the three different regions,
which results in an energy quantization condition.
This method can be straightforwardly extended to the more complex
situation studied here by employing the general solution in Sec.~\ref{sec2}
for the central strip.  

Before turning to results, we briefly summarize
the parameter values chosen 
in numerical calculations. We take a magnetic field value $B=0.2$~T,
and the waveguide width is $2L= \sqrt{8} \ell_B\approx 40$~nm.
The strain-induced parameters in the
central strip are taken as ${\cal A}_x= - 16 \mu$m$^{-1}$ and
$V= - 20$~meV.  
These values have been estimated for a folded setup,\cite{rainis}
where $V$ comes from the deformation potential.
We consider two different parameter choices for the SOI couplings:
Set (A) has $\Delta=13$~meV and $\lambda=3$~meV, corresponding to
the QSH phase.  For set (B), we exchange both values, i.e., 
$\Delta=3$~meV and $\lambda=13$~meV.

\subsection{Antiparallel case: Snake orbit}

\begin{figure}
\includegraphics[width=0.45\textwidth]{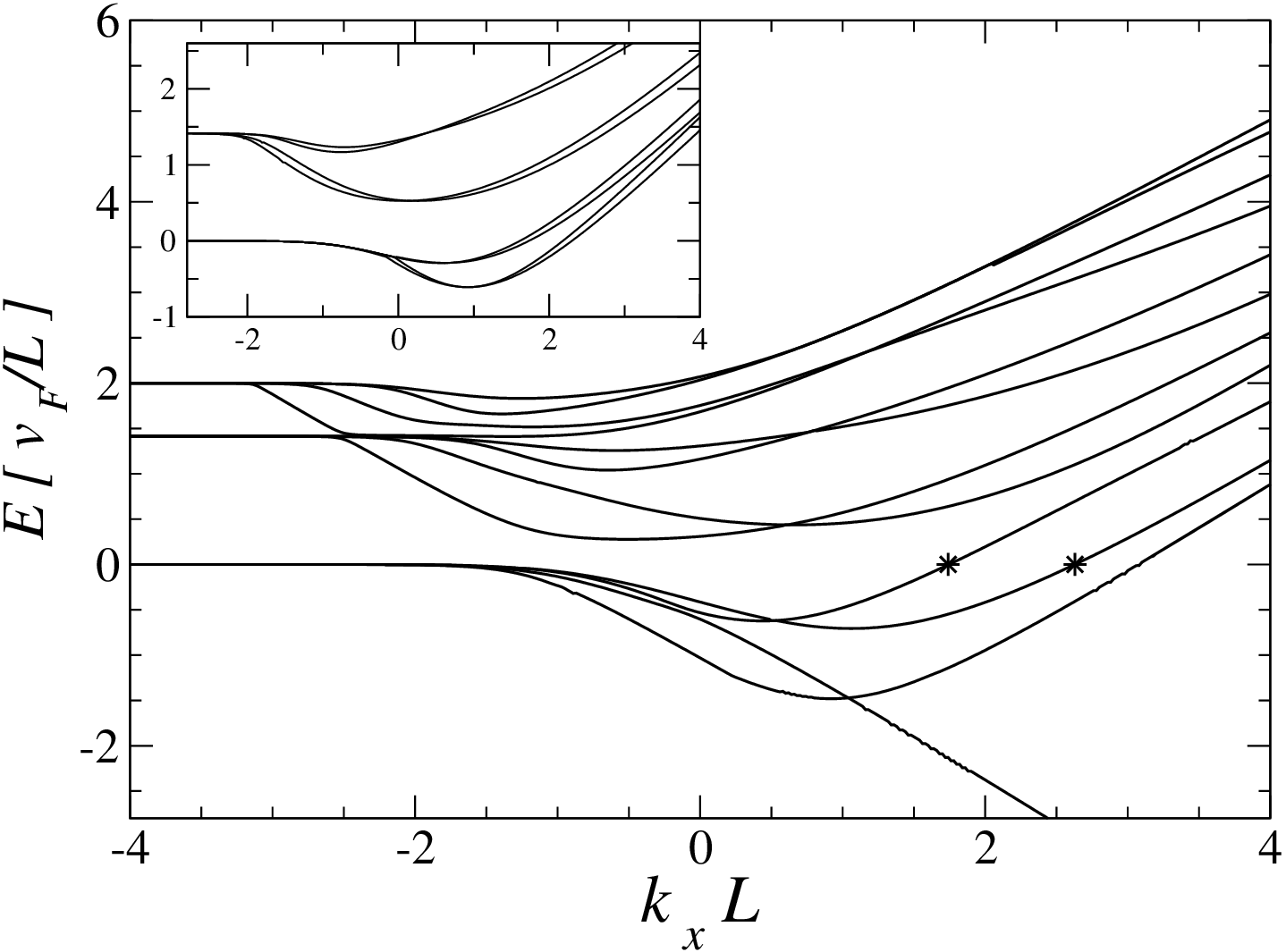}
\caption{\label{fig4}    
Dispersion relation of the lowest few energy branches for 
a strained magnetic waveguide with $\epsilon=-1$ 
and SOI in the central strip of width $2L$. Energies are given in units of 
$\hbar v_F/L$.  The main panel is for parameter set (B). 
The stars refer to the states further studied in Fig.~\ref{fig5}.
Inset: Same for set (A).  (See main text for details.)  }
\end{figure}

Let us first discuss the $\epsilon=-1$ configuration, where the
magnetic field $B_z$ differs in sign in the regions $y<-L$ and $y>L$.
The dispersion relation of typical low-energy 1D waveguide modes 
is shown in Fig.~\ref{fig4}.  For $k_x\to -\infty$ the centers of the quantum states 
are located deep in the left and right magnetic regions, far from the waveguide.
Thus one has doubly-degenerate dispersionless ``bulk'' Landau states.
With increasing $k_x$ these states are seen to split up.
The dominant splitting, which is already present for $\Delta=\lambda=0$,
comes from the splitting of symmetric and anti-symmetric linear combinations 
of the Landau states for $y<-L$ and $y>L$ with increasing overlap in the 
waveguide region.\cite{ghosh}
Asymptotically, the dispersion relation of all positive-energy
snake states is $E(k_x\to +\infty) \simeq\hbar v_F k_x$.\cite{ghosh}
For intermediate $k_x$ and $(\Delta,\lambda)\ne 0$, however,
we get \textit{spin-split snake states} out of the previously 
spin-degenerate states.  
The spin splitting is mainly caused by the Rashba coupling $\lambda$
and disappears for $\lambda\to 0$, cf.~the inset of Fig.~\ref{fig4}.

\begin{figure}
\vspace{2cm}
\includegraphics[width=0.45\textwidth]{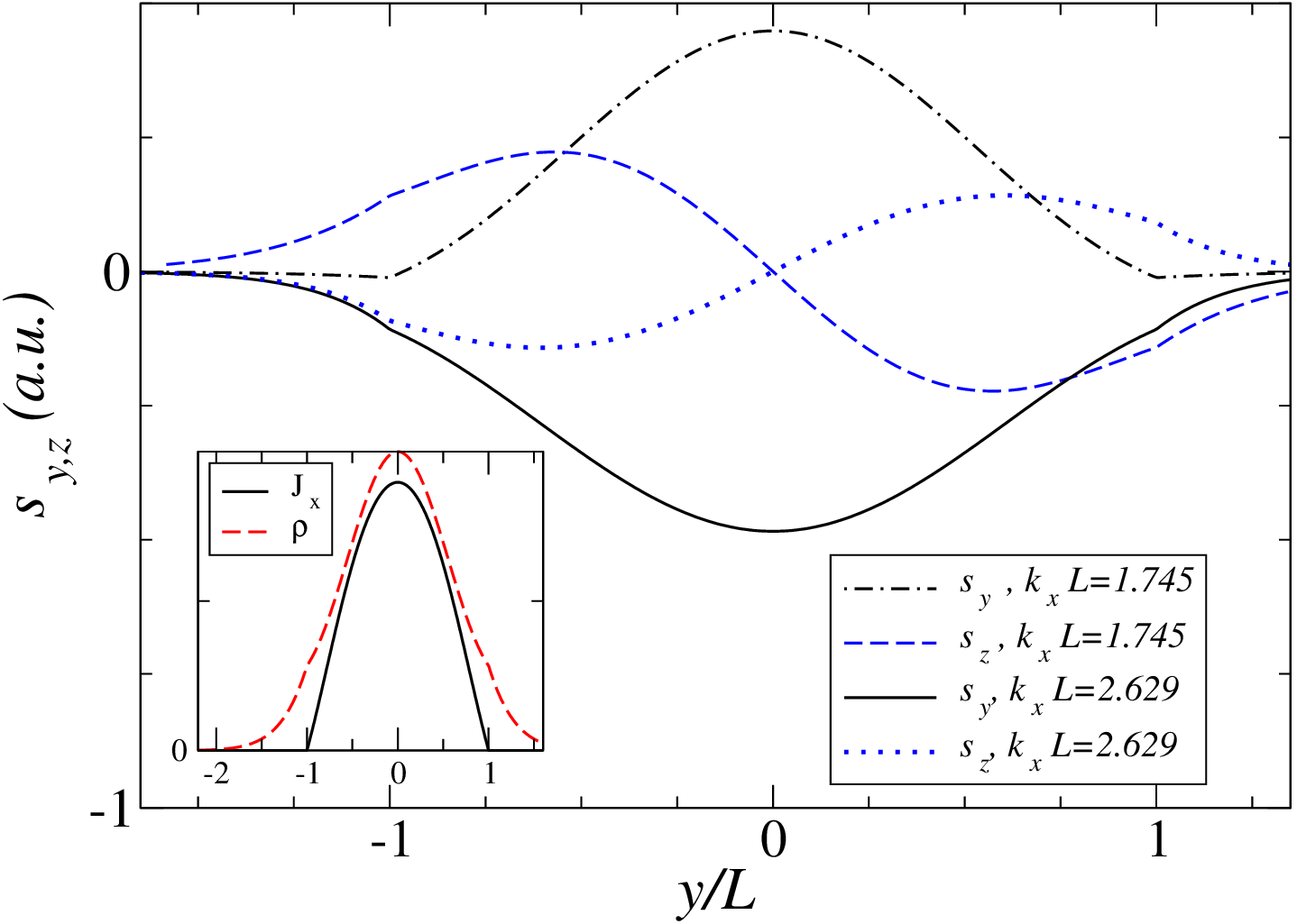}
\caption{\label{fig5}  (Color online)   
Spin density profile $S_{y,z}$ (in arbitrary units) 
vs $y/L$ for the two $E=0$ 
right-moving states indicated by stars 
in the main panel of Fig.~\ref{fig4}.  The left star corresponds to 
$k_xL=1.745$, the right star to $k_x L=2.629$. 
Inset: Particle density, $\rho$, and current
density, $J_x$ (which is the only non-vanishing component), 
in arbitrary units vs $y/L$.  We show the result only for $k_x L=1.745$,
since $k_x L=2.629$ yields practically the same. }
\end{figure}

The zero-energy bulk Landau state (for $k_x\to -\infty$)
shows rich and interesting behavior in this setup.  
While for $k_x\to +\infty$, we 
expect one pair of snake states with positive slope and one pair with negative
slope, for the studied parameter set and range of $k_x L$, there is just one 
state with negative slope while three branches first move down
and then have a positive slope.  Accordingly, at the Dirac point ($E=0$),
Fig.~\ref{fig4} shows that there are three right-movers with different
Fermi momenta and different spin texture.  Two of those states
are indicated by stars (*) in the main panel of Fig.~\ref{fig4} and
their local spin texture is shown in Fig.~\ref{fig5}.  Evidently, they
are mainly localized inside the waveguide and have antiparallel spin
polarization. We find spin densities with $S_x=0$ for both states.
For the Rashba-dominated situation in Fig.~\ref{fig5}, spin is
polarized perpendicular to the current direction and has a rather complex
spatial profile.

\subsection{Parallel configuration}

\begin{figure}
\vspace{2cm}
\includegraphics[width=0.45\textwidth]{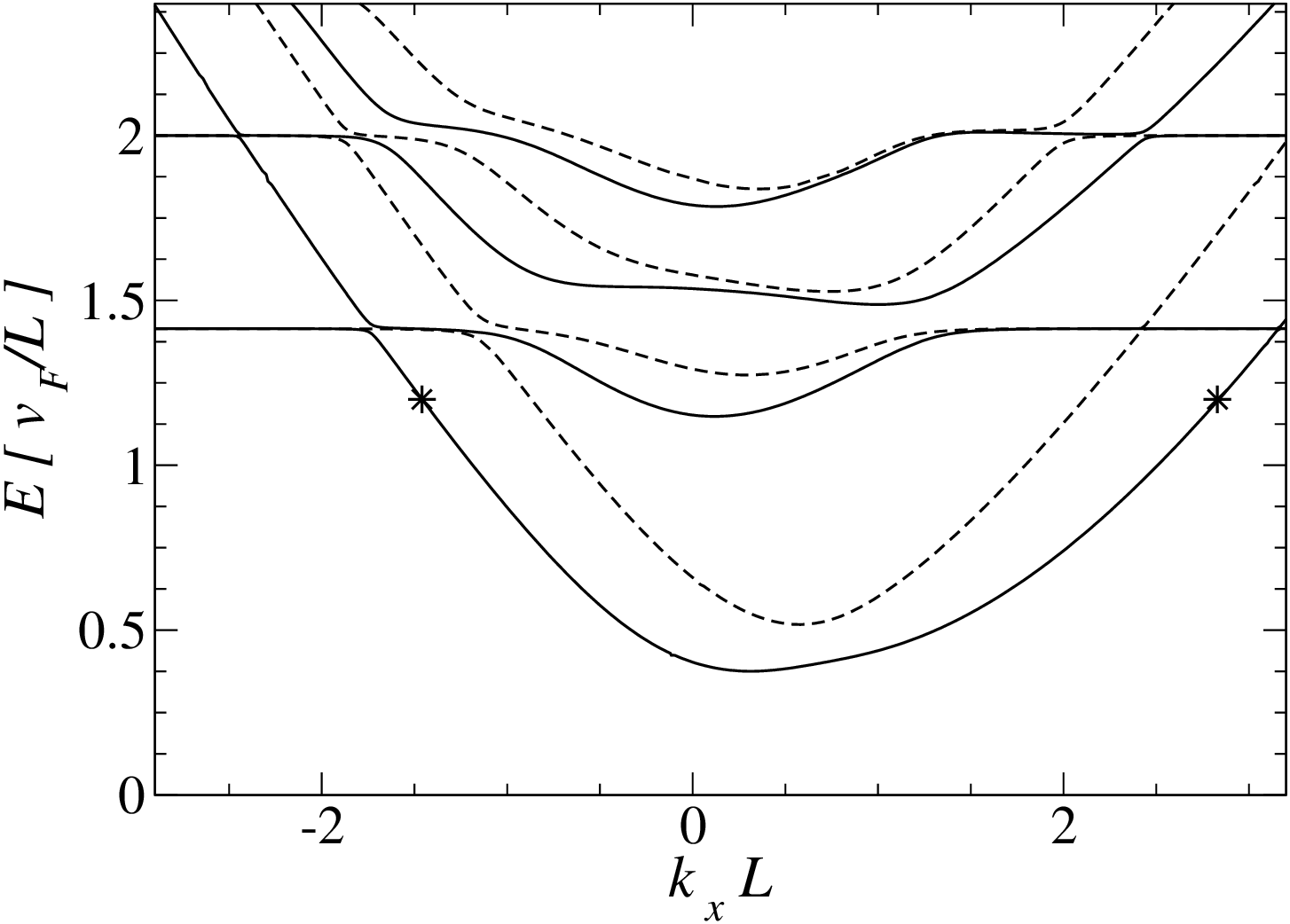}
\caption{\label{fig6}   Same as Fig.~\ref{fig4} but
for the setup with $\epsilon=+1$ and parameter set (B).  
Solid and dashed curves are for better visibility only.
The two states indicated by stars are studied in Fig.~\ref{fig7}.
}
\end{figure}

\begin{figure}
\vspace{2cm}
\includegraphics[width=0.45\textwidth]{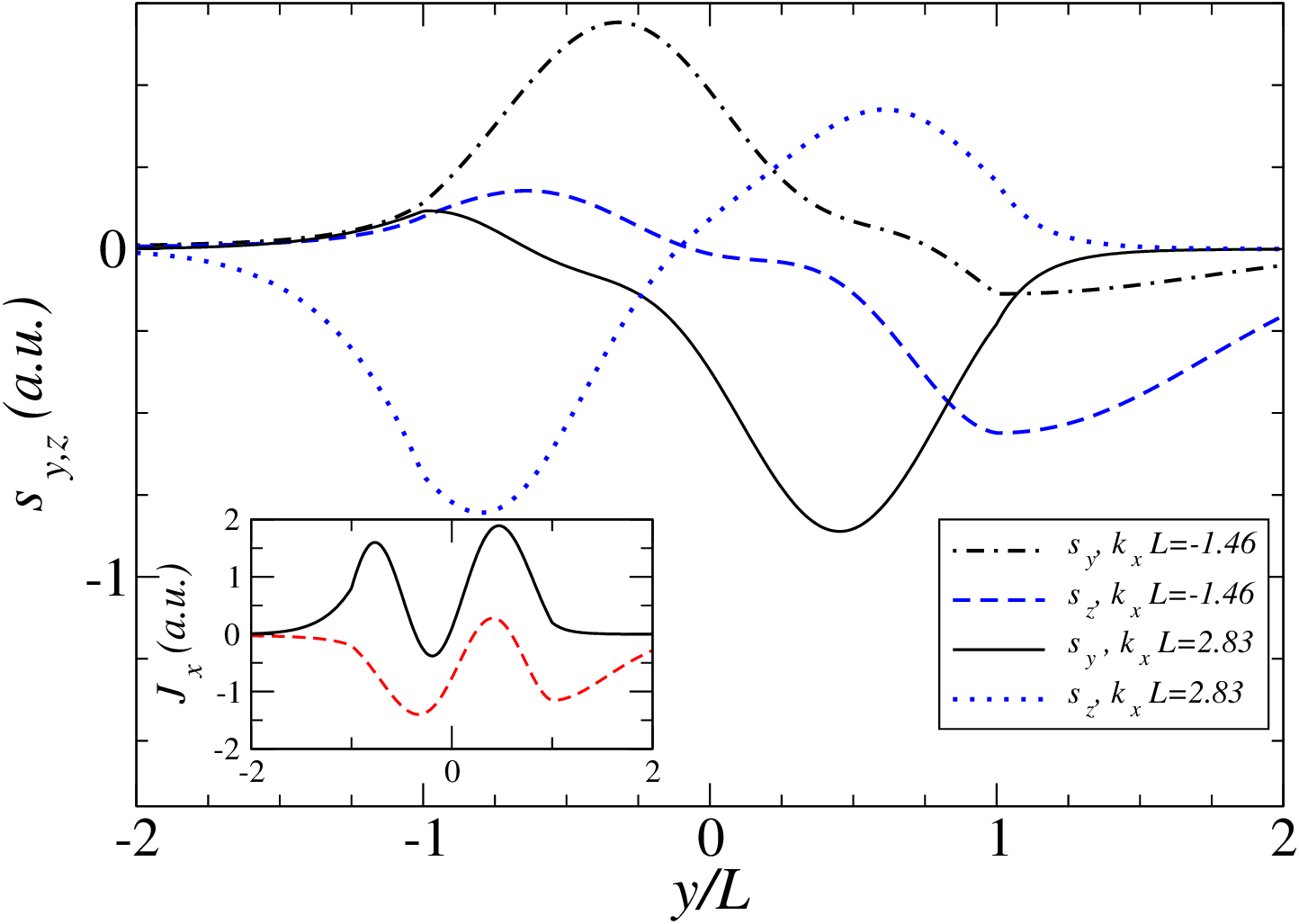}
\caption{\label{fig7}  (Color online)   
Spin density $S_{y,z}$ (in arbitrary units) vs $y/L$ 
for the two states indicated by stars in Fig.~\ref{fig6}.
The left (right) star corresponds to a left- (right-)mover with 
$k_x L=-1.46$  ($k_x L =2.83$).  Note that the spin polarizations
of both states are approximately antiparallel.  Inset: Particle 
current profile $J_x$ (in arbitrary units) vs $y/L$
for both states. Black solid curve: $k_x L=2.83$. Dashed red curve: 
$k_x L=-1.46$.  }
\end{figure}

Next we come to the $\epsilon=+1$ configuration, where the 
magnetic field is $+B$ for $|y|>L$ and $-B$ for $|y|<L$.
One therefore expects two counterpropagating snake states in the
$x$-direction localized around $y=\pm L$.  The corresponding 
spectrum is shown in Fig.~\ref{fig6}.  We focus on parameter set (B),
since for set (A), the spin splitting is minimal and less interesting.
The spectrum consists of two qualitatively different states, namely
states of bulk Landau character for large $|k_x| L$, 
and a set of propagating waveguide modes.\cite{ghosh}
The spectral asymmetry seen in Fig.~\ref{fig6} 
for all propagating modes, $E(-k_x)\ne E(k_x)$, 
is caused by the strain (${\cal A}_x$)-induced shift of $k_x$.
Such a spectral asymmetry may give rise to interesting 
chirality and magnetoasymmetry effects.\cite{tsvelik} 
The spin texture is shown in Fig.~\ref{fig7}
for a pair of right- and left-moving states with $E=1.2 \hbar v_F/L$,
cf.~the stars in Fig.~\ref{fig6}.
We observe from the main panel in Fig.~\ref{fig7} that the spin polarization
of both states is approximately antiparallel.  Because of their spatial
separation and the opposite spin direction, elastic disorder 
backscattering between
these counterpropagating snake modes should be very strongly
suppressed.  The inset of Fig.~\ref{fig7} shows the current density 
profile across the waveguide.  Although the profile is quite complex,
we observe that the current  has opposite sign for both modes.

\section{Concluding remarks}
\label{sec6}

In this work, we have studied the magnetoelectronic 
properties of monolayer graphene in the presence of strong 
intrinsic and Rashba-type  spin-orbit couplings.
According to a recent proposal,\cite{franz} large intrinsic
couplings may be realized by suitable adatom deposition on graphene.
We have presented an exact solution for the Landau level states for 
arbitrary SOI parameters. When the intrinsic SOI dominates, by increasing 
the magnetic field, we predict a quantum phase transition from the quantum 
spin Hall phase to a helical quantum Hall phase at the Dirac point.  
In both phases, one has spin-filtered edge states but with opposite
spin current direction. Thus the transition could be detected by 
measuring the spin current either in a transport experiment (e.g., along
the lines of Ref. \onlinecite{tombros}) or via a magneto-optical experiment.

In inhomogeneous magnetic fields, especially
when also strain-induced pseudo-magnetic fields are present,
interesting waveguides can be envisioned.  Such setups allow for
snake states, where spin-orbit couplings result in a spin splitting.
In a double-snake setup, there is a pair of counterpropagating 
snake states that carry (approximately) opposite spin polarization.
This implies that scattering by elastic impurities is drastically
suppressed. The resulting spin textures can in principle be detected  
by spin resolved ARPES (see, e.g., Refs. \onlinecite{dedkov}
and \onlinecite{bostwick}) 
or spin-polarized STM measurements.

We hope that our predictions can soon be tested experimentally.

\acknowledgments
We acknowledge financial support by the DFG programs SPP 1459 and SFB TR 12.

\appendix
\section{Derivation of the eigenstates} \label{appa}

Here we provide some details concerning the derivation
of Eq.~(\ref{sol1}); the notation below is explained in Sec.~\ref{sec2}.
First, additional relations like Eq.~\eqref{parbol} can be stated,
\begin{eqnarray*}
a D_p(-\eta) &=& - p D_{p-1}(-\eta), \quad a^\dagger D_p(-\eta) = 
-D_{p+1}(-\eta),\\ 
a D_p(i\eta) &=&- i D_{p+1}(i\eta), \quad a^\dagger D_p(i\eta) = 
-ipD_{p-1}(i\eta).
\end{eqnarray*}
We wish to construct the solution $\phi=(\phi_1,\phi_2,\phi_3,\phi_4)^T$
satisfying Eq.~(\ref{schr}), 
\[
\left( \begin{array}{cccc} \nu_- & a & 0 & 0 \\
a^\dagger & \nu_+ & i\lambda & 0\\
0 & -i\lambda & \mu_+ &  a\\
0 & 0  &  a^\dagger & \mu_-
\end{array}\right) \left( \begin{array}{c}
\phi_1 \\ \phi_2\\ \phi_3 \\ \phi_4 \end{array}\right) = 0.
\]
We here show only the case $\epsilon=+1$ near the $K$ point; all
other cases follow analogously.
Solving the first and last equations for $\phi_1$ and $\phi_4$,
respectively, we find
\[
\phi_1 = -\frac{1}{\nu_-} a\phi_2,\quad \phi_4=-\frac{1}{\mu_-}a^\dagger \phi_3.
\]
For $\mu_-=0$, one has the solution (\ref{p-1}) instead, while 
for $\nu_-=0$ there are no solutions.
The second and third equations then yield two coupled second-order 
ordinary differential equations for $\phi_2$ and $\phi_3$,
\begin{eqnarray*}
(a^\dagger a -\nu)\phi_2 -i\lambda\nu_-\phi_3 &=& 0,\\ 
i\lambda\mu_-\phi_2+ (a^\dagger a+1 -\mu)\phi_3 &=& 0.
\end{eqnarray*}
Solving for $\phi_3$ yields
\[
\phi_3= \frac{1}{i\lambda\nu_-} (a^\dagger a-\nu)\phi_2,
\]
and we thus arrive at the equation
\[
{\cal D}\phi_2\equiv 
\left[ (a^\dagger a+1-\mu)(a^\dagger a-\nu) -\lambda\mu_-\nu_-\right]\phi_2=0.
\]
Since the operator ${\cal D}$ commutes with the ``number operator'' 
$a^\dagger a$, the sought
solutions for $\phi_2$ span the kernel of ${\cal D}$  where
the $\phi_2$  are eigenstates of $a^\dagger a$,
\[
a^\dagger a \ \phi_{2,p}(\eta)= p \ \phi_{2,p}(\eta).
\]
This leads to an algebraic equation for the eigenvalue $p$,
\[
(p+1-\mu)(p-\nu)=\lambda^2 \mu_-\nu_-,
\]
which implies the two solutions in Eq.~\eqref{pdef}.
With $a^\dagger a=(\eta/2)^2-1/2-d^2/d\eta^2$, the eigenvalue equation
for $\phi_2$ is just the differential equation of the parabolic
cylinder functions,\cite{abram}
\[
\left( \frac{d^2}{d\eta^2} +p+\frac12 - \frac{\eta^2}{4}\right) \phi_2(\eta)=0,
\]
which has the four (linearly dependent) solutions $\{ D_p(\eta),
D_p(-\eta),D_{-p-1}(i\eta), D_{-p-1}(-i\eta)\}$.  Given the solution
for $\phi_2$, all other components in $\phi$ follow by using the 
recurrence relations of the $D_p$ functions, see, e.g., Eq.~\eqref{parbol}.
After straightforward but lengthy algebra, we obtain the four 
solutions (also quoted for $\epsilon=-1$)
\begin{widetext}
\begin{eqnarray*}
\phi_{\epsilon=+1,p} &=&\left( \begin{array}{c} 
- p D_{p-1}(\eta) \\ \nu_- D_p(\eta)\\
\frac{i(\nu-p)}{\lambda} D_p(\eta)\\ \frac{-i(\nu-p)}{\lambda \mu_-} 
D_{p+1}(\eta) \end{array} \right),\quad
\left( \begin{array}{c} p D_{p-1}(-\eta) \\ \nu_- D_p(-\eta)\\
\frac{i(\nu-p)}{\lambda} D_p(-\eta)\\ \frac{i(\nu-p)}{\lambda \mu_-} 
D_{p+1}(-\eta) \end{array} \right) , \\
\psi_{\epsilon=+1,p} & = & \left( \begin{array}{c}
- i D_{-p}(-i\eta) \\ \nu_- D_{-p-1}(-i\eta)\\ 
\frac{i(\nu-p)}{\lambda} D_{-p-1}(-i\eta)\\
\frac{-(\nu-p)(p+1)}{\lambda \mu_-} D_{-p-2}(-i\eta) \end{array} \right), \quad
\left( \begin{array}{c} -i D_{-p}(i\eta) \\ \nu_- D_{-p-1}(i\eta)\\
\frac{i(\nu-p)}{\lambda} D_{-p-1}(i\eta)\\ \frac{(\nu-p)(p+1)}
{\lambda \mu_-} D_{-p-2}(i\eta) \end{array}
\right),\\
\phi_{\epsilon=-1,p} & =& 
\left( \begin{array}{c} 
-D_{p+1}(-\eta) \\ \mu_- D_p(-\eta)  \\
\frac{i(\mu-p-1)}{\lambda} D_p(-\eta)\\
\frac{-i(\mu-p-1)}{\lambda \nu_-} D_{p-1}(\eta)
\end{array} \right),\quad \left( \begin{array}{c} D_{p+1}(\eta) \\ 
\mu_- D_p(\eta)\\ \frac{i(\mu-p-1)}{\lambda} D_p(\eta)\\
\frac{i(\mu-p-1)}{\lambda \nu_-} D_{p-1}(\eta) \end{array} \right), \\
\psi_{\epsilon=-1,p} & =&\left( \begin{array}{c} - i (p+1)D_{-p}(-i\eta) \\
\mu_- D_{-p-1}(-i\eta)\\
\frac{i(\mu-p-1)}{\lambda} D_{-p-1}(-i\eta)\\
\frac{(\mu-p-1)}{\lambda \nu_-} D_{-p-2}(-i\eta)
\end{array} \right), \quad
\left( \begin{array}{c} i (p+1)D_{-p}(i\eta) \\ \mu_- D_{-p-1}(i\eta)\\
\frac{i(\mu-p-1)}{\lambda} D_{-p-1}(i\eta)\\
-\frac{(\mu-p-1)}{\lambda \nu_-} D_{-p-2}(i\eta) \end{array} \right).
\end{eqnarray*}
\end{widetext}
For a given energy $E$, Eq.~\eqref{schr} admits precisely 
four linearly independent solutions for $\phi$.  However, Eq.~\eqref{pdef}
implies two possible values for $p$, i.e., we have the freedom to choose
just two out of the four quoted eigenstates (for given $\epsilon$) and
then allow both values of $p$ in Eq.~\eqref{pdef}.   
Our conventions for these two basis states are 
specified in Eqs.~\eqref{sol1} and \eqref{sol2} in the main text.
Thereby we have obtained all possible solutions to Eq.~\eqref{schr}.


\begin{thebibliography}{10}
\bibitem{footnote1}
A different representation for ${\cal T}$ has been given in 
Ref.~\onlinecite{beenakker} because of a different 
arrangement of the sublattice components in the spinor [Eq.~\eqref{spinor}].
\bibitem{foot5}
For notational convenience, we shift $p+1\to p$ for $s=\downarrow$
in the discussion of the purely intrinsic SOI.
\bibitem{foot2}
We note that we made a wrong statement in Ref.~\onlinecite{ssc} in that
direction: For the case $j<0$ on page 3 therein, we stated that for $a'=0$
there are two normalizable states with $E=\pm  |M|$ which for $M\to 0$
coalesce into a single zero-energy Landau level. However, only the state
$E=-|M|$ is allowed, and the other one is not normalizable.
\end{thebibliography}
\end{document}